%% file: ew_lensed_cbc.tex
\documentclass[iop]{emulateapj}

\usepackage{physics}
\usepackage{amsmath,color,bm}
\usepackage{booktabs}
\usepackage[varg]{txfonts}
\usepackage{multirow}

\usepackage{natbib} 

\makeatother

\usepackage{nameref}
\usepackage{graphicx}
\usepackage{graphics}
\usepackage[space]{grffile}
\usepackage{url}
\usepackage[utf8]{inputenc}
\usepackage{fancyref}
\usepackage{hyperref}
\usepackage{gensymb}
\usepackage[dvipsnames]{xcolor}

\hypersetup{
colorlinks=true,
linkcolor=BlueViolet,
urlcolor=BlueViolet,
citecolor = BlueViolet
}

\newcommand\blfootnote[1]{%
  \begingroup
  \renewcommand\thefootnote{}\footnote{#1}%
  \addtocounter{footnote}{-1}%
  \endgroup
}


\begin{document}
\input{title.tex}

\input{abstract.tex}

\section{Background}\label{sec:introduction}
	\input{introduction.tex}
\section{Method}\label{sec:method}
\input{method.tex}

\section{Results}\label{sec:results}
	\input{results.tex}

\section{Discussion}\label{sec:conclusion}
	\input{conclusion.tex}

\section{Acknowledgements}
We are very grateful to M. K. Haris for providing his code, based on \citep{haris2018}, portions of which were used to generate the time delay distributions. We also thank Otto Hannuksela for a careful review of this work, and Tejaswi Venumadhav for useful discussions. All computations were performed on the IUCAA computing cluster Sarathi, and the ICTS computing cluster Alice. This work makes use of \texttt{NumPy} \citep{vanderWalt:2011bqk}, \texttt{SciPy} \citep{Virtanen:2019joe}, \texttt{astropy} \citep{2013A&A...558A..33A, 2018AJ....156..123A}, \texttt{Matplotlib} \citep{Hunter:2007}, \texttt{jupyter} \citep{jupyter}, \texttt{pandas} \citep{mckinney-proc-scipy-2010}, \texttt{pycbc} \citep{pycbc} software packages.

\bibliographystyle{aasjournal}
\bibliography{references.bib}

\onecolumngrid
\vspace{-1cm}
\begin{appendix}

\input{appendix.tex}

\end{appendix}

\end{document}

%% file: title.tex
\title[]{Gear-up for the Action Replay: Leveraging Lensing for Enhanced Gravitational-Wave Early-Warning}


\author{Sourabh Magare $\dagger$ \href{https://orcid.org/0000-0002-0870-2993}{\includegraphics[scale=0.03]{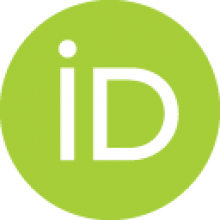}},$^{1}$   Shasvath J. Kapadia $\ddagger$ \href{https://orcid.org/0000-0001-5318-1253}{\includegraphics[scale=0.03]{orcid-ID.png}},$^{1,2}$ Anupreeta More \href{https://orcid.org/0000-0001-7714-7076}{\includegraphics[scale=0.03]{orcid-ID.png}},$^{1,3}$ Mukesh Kumar Singh \href{https://orcid.org/0000-0001-8081-4888}{\includegraphics[scale=0.03]{orcid-ID.png}},$^{2}$  Parameswaran Ajith \href{https://orcid.org/0000-0001-7519-2439}{\includegraphics[scale=0.03]{orcid-ID.png}}, $^{2,4}$ and \\ A. N. Ramprakash \href{https://orcid.org/0000-0001-5707-4965}{\includegraphics[scale=0.03]{orcid-ID.png}} $^{1}$}

\address{$^{1}$ The Inter-University Centre for Astronomy and Astrophysics, Post Bag 4, Ganeshkhind, Pune 411007, India \\
$^{2}$ International Centre for Theoretical Sciences, Tata Institute of Fundamental Research, Bangalore 560089, India\\
$^{3}$ Kavli Institute for the Physics and Mathematics of the Universe (IPMU), 5-1-5 Kashiwanoha, Kashiwa-shi, Chiba 277-8583, Japan \\
$^{4}$ Canadian Institute for Advanced Research, CIFAR Azrieli Global Scholar, MaRS Centre, West Tower, 661 University Ave, Toronto, ON M5G 1M1, Canada
}


%% file: abstract.tex
\begin{abstract}
Pre-merger gravitational-wave (GW) sky-localisation of binary neutron star (BNS) and neutron star black hole (NSBH) coalescence events, would enable telescopes to capture precursors and electromagnetic (EM) emissions around the time of the merger.  We propose a novel astrophysical scenario that could provide early-warning times of hours to days before coalescence with sub-arcsecond localisation, provided that these events are gravitationally lensed. The key idea is that if the BNS/NSBH is lensed, then so must the host galaxy identified via the EM counterpart. From the angular separation of the lensed host galaxy images, as well as its redshift and the (foreground) lens redshift, we demonstrate that we can predict the time delays assuming a standard lens model. Encouraged by the non-trivial upper limits on the detection rates of lensed BNS/NSBH mergers that we estimate for upcoming observing runs of the LIGO-Virgo-Kagra and third generation networks, we assess the feasibility and benefits of our method. To that end, we study the effect of limited angular resolution of the telescopes on our ability to predict the time delays. We find that with an angular resolution of $0.05''$, we can predict time delays of $> 1$ day with $1\sigma$ error-bar of $\mathcal{O}$(hours) at best. We also construct realistic time delay distributions of detectable lensed BNSs/NSBHs to forecast the early-warning times we might expect in the observing scenarios we consider.
\end{abstract}

\keywords{Gravitational Waves, Gravitational Lensing, Multi-Messenger Astronomy}

%% file: introduction.tex
Multi-messenger astronomy involving gravitational-waves (GWs) arrived with the GW detection of binary neutron star (BNS) merger GW170817 \citep{GW170817-DETECTION} by the LIGO--Virgo \citep{advligo, advvirgo} network of detectors, and the extensive electromagnetic (EM) follow-up by telescopes worldwide \citep{GW170817-MMA}.\blfootnote{$\dagger$sourabh.magare@iucaa.in}\blfootnote{$\ddagger$shasvath.kapadia@iucaa.in} 


While the success of GW170817 cannot be underplayed \citep[see, e.g.][for a review]{stratta2022}, there was nevertheless a delay between the occurrence of the corresponding GW event, and the subsequent EM follow-up \citep{GW170817-MMA}. As a result, certain pre-merger \citep[see, e.g.][]{tsang2012, Most2020}, merger and post-merger EM emissions \citep[see, e.g.][]{siegel2016-1, siegel2016-2} may have been missed. A possible way to capture these emissions is to alert telescopes sufficiently early in the inspiral of the binary so that they can scan a well localised patch of the sky prior to the merger.


Current matched-filter-based GW early-warning methods rely on accumulating adequate signal-to-noise ratio (S/N) before the merger of the binary to allow for its prompt detection and sky-localisation \citep{CannonEW}. BNS events are especially amenable to this method; their smaller masses -- compared to binary black holes (BBHs) and neutron-star black hole (NSBH) binaries -- allow them to spend a relatively longer time in the frequency band of the LIGO--Virgo detectors. 

Nevertheless, the typical early-warning times that can be expected for BNSs in LIGO--Virgo's (and Kagra's \citep{KAGRA:2020cvd, KAGRA:2020tym}) next observing run (O4) is $\sim 10$~seconds \citep{Sachdev:2020lfd, Magee:2021xdx, Nitz:2020vym}, corresponding to a localisation sky area of $O(100)$~sq.~deg. Using an approximate triangulation-based localisation method \citep{Fairhurst1, Fairhurst2}, we estimate that this early-warning time increases to $\sim 30$~seconds in O5 \footnote{See also \citep{Nitz:2020vym} for more detailed pre-merger localisation estimates in O5.} \citep{KAGRA:2013rdx}, and $\sim 1$~min in the Voyager \citep{adhikari2018} scenario. Such early-warning times might not be adequate for telescopes to capture precursors, or even EM emissions surrounding the merger.

Prospects are even less optimistic for NSBH systems, a fraction of which are expected to be EM-Bright \citep{FoucartEMB}, depending on the equation of state of the NS and the spin of the BH. Their heavier masses make them spend an even smaller time in-band, which makes pre-merger alerts for their coalescence more challenging. The in-band time can effectively be stretched, for certain NSBH configurations, by the inclusion of higher-harmonics of GW radiation in real-time searches \citep{Kapadia:2020kss,Singh:2020lwx,Singh:2022tlh}. But this provides an improvement of a factor of a few at best in early-warning time \footnote{If the NSBH is precessing, an additional $\mathcal{O}(10)$ sec of early-warning time can be purchased \citep{Tsutsui:2021izf}.} -- which is still likely to be insufficient to capture merger emissions and precursors.

\section{Early-Warning with Lensing}
In this work, we envisage a novel astrophysical scenario that could enable early-warning times of hours to days, with a sub-arcsecond sky localisation, involving gravitational lensing. The method relies on a simple idea: if the BNS or NSBH is strongly lensed (say by a galaxy or a cluster), then so must the host-galaxy. Leveraging the properties of the host-galaxy's lensed images, early-warning can be enhanced drastically, as follows. 

The EM-counterpart would enable the identification of the host galaxy during EM follow-up of the BNS/NSBH merger. Image analysis of the host-galaxy would then indicate if it is lensed. If required, subsequent observation of the host-galaxy by high-resolution telescopes would provide precise astrometry. These, in conjunction with the redshift measurements of the host-galaxy and the foreground lens galaxy, could be used to predict the time-delay between multiple lensed images/events. This would effectively act as the early-warning-time for the occurrence of the ``action-replay'' of the BNS/NSBH merger due to lensing. The idea is similar in spirit to Supernova (SN) ``Refsdal'' \citep[see, e.g.][]{Kelly:2015xvu}, where observed lensed images of the transient SN enabled a prediction of the occurrence of the next lensed SN image in the future.

\section{Detection rate of lensed BNS and NSBH mergers}

We estimate the rates of detectable lensed BNS and NSBH mergers, following standard prescriptions in the literature (see Appendix~\ref{app:rate} for details). We consider the O4, O5 and Voyager observing scenarios, as well as the third generation (3G) detector network \citep[see, e.g.][]{Hall2019}. The results are tabulated in Table~\ref{tab:rate}.

Note the considerable uncertainty in the estimated rates. This primarily stems from the fact that no lensed BNSs/NSBHs have been detected so far, and barely a handful of corresponding unlensed systems have been observed with GWs.

\begin{table}
\centering
\begin{tabular}{|l|c|c|c|c|c} \hline
Obs. & Lensed BNS $[\mathrm{yr}^{-1}]$ & Lensed NSBH $[\mathrm{yr}^{-1}]$ \\\hline
O4 & $0 - 3.0  \times 10^{-1}$ & $1.0 \times 10^{-3} - 2.0 \times 10^{-1}$ \\\hline 
O5 & $5.0 \times 10^{-3} - 1.0 \times 10^1$ & $2.0 \times 10^{-2} - 3.4$ \\\hline
Voyager & $3.0 \times 10^{-1} - 6.2 \times 10^2$ & $7.0 \times 10^{-1} - 1.8 \times 10^2$ \\\hline
3G & $1.2 \times 10^1 - 3.5 \times 10^4$ & $1.2 \times 10^1 - 4.3 \times 10^3$ \\\hline
\end{tabular}
\caption{Rates of detectable lensed BNS and NSBH events, for various observing scenarios. Only galaxy lenses are considered, which is expected to be a good approximation given the relative rarity of clusters. See Appendix~\ref{app:rate} for additional details on the assumptions used to estimate the rates.}
\label{tab:rate}
\end{table}


The estimated rates suggest that seeing a lensed BNS or NSBH in O4 is unlikely, although not abysmally small assuming the optimistic upper limit. Seeing such an event in O5 appears much more likely relative to O4. Voyager seems to be the first observing scenario that will almost certainly see a lensed BNS or NSBH, and the 3G scenario guarantees it.

That said, ultimately, the detection of lensed BNSs/NSBHs will follow Poisson statistics, and a Poisson fluctuation may enable observations of such events even in O4. Indeed, being ready for such an event, in case one should present itself, could have dramatic scientific payoffs due to the drastic enhancement of early-warning times. 

This motivates us to assess the practicability and benefits of our method. To that end, we first characterize and quantify uncertainties on the time delay predictions arising from various systematics. In this work, we focus on the astrometric uncertainties when analysing the optical images of the lensed host galaxy. We then provide prospective time delay distributions of detectable BNS/NSBH lensed by galaxy-scale lenses, to gauge the kind of early-warning times we might expect in various observing scenarios. We assume standard cosmology \citep{Planck2018} throughout.

%% file: method.tex
GWs, like light, will have their trajectories deviated if they encounter large assemblages of matter \citep{gunn1967, Wang:1996as}. 
Thus, galaxy or cluster scale lenses can produce two or more non-overlapping copies of GWs emanated by stellar-mass compact binary coalescences (CBCs). The copies will have identical phase evolution, but differing amplitudes \citep[see, e.g.][]{haris2018, Dai2020}. Searches for such (and other) lensing signatures in GW data have been conducted. No confirmed detection has so far been reported \citep{LIGOScientific:2021izm, Dai2020, Hannuksela:2019kle} \footnote{Alternatively, it has been suggested that a significant fraction of LIGO-Virgo's BBH detections correspond to magnified lensed events, which is why the inferred masses of the components of the BBHs are larger in comparison to those observed from EM observations \citep{broadhurst2018, broadhurst2022}}.  

Here, we describe the basics of the lens equation used to predict time delays between images. We also provide the framework for mock sample generation of lens systems to construct time delay distributions of detectable lensed events in various observing scenarios.  We adopt the singular isothermal ellipsoid (SIE) model appropriate for isolated galaxy lenses \citep{koopmans2009}. 
We work in the geometric optics limit, valid for GWs emanated by stellar-mass CBCs that encounter galaxy lenses. We use the thin lens approximation because the size ($\sim$ kpc) of the lens is significantly smaller than the distances separating the source from the lens ($\sim$ Mpc/Gpc), and the lens from the earth ($\sim$ Mpc/Gpc). 

\subsection{The lens equation}

Let $\vec{y} = (y_1, y_2),~\vec{x} = (x_1, x_2)$ be the position vectors of the source in the source plane and its image(s) in the lens plane. The standard problem in lensing consists of acquiring the image s $\vec{x}$, from the source position $\vec{y}$ and the lens potential \citep[e.g.,][]{schneider1992, dodelson2017}:
\begin{equation}\label{eq:lens}
\vec{y} = \vec{x} - \vec{\alpha}(\vec{x})
\end{equation}    
where the deflection angle $\vec{\alpha} = \nabla\psi(\vec{x})$ captures the influence of the effective lens potential $\psi$. This potential, in turn, can be acquired from the surface mass density profile $\Sigma(\vec{x})$:
\begin{equation}\label{eq:potential}
\psi = \frac{1}{\pi}\int\kappa(\vec{u})\ln|\vec{x} - \vec{u}|d\vec{u}
\end{equation}
where $\kappa \equiv \Sigma/\Sigma_c$ is the normalized mass surface density profile, $\Sigma_c \equiv (c^2/4\pi G) (D_s /D_l D_{ls})$ is the critical mass density, and $D_l, D_s, D_{ls}$ are the angular diameter distances between the earth and lens, the earth and source and the lens and the source, respectively.

\begin{figure}
	\centering 
	\includegraphics[width=\columnwidth]{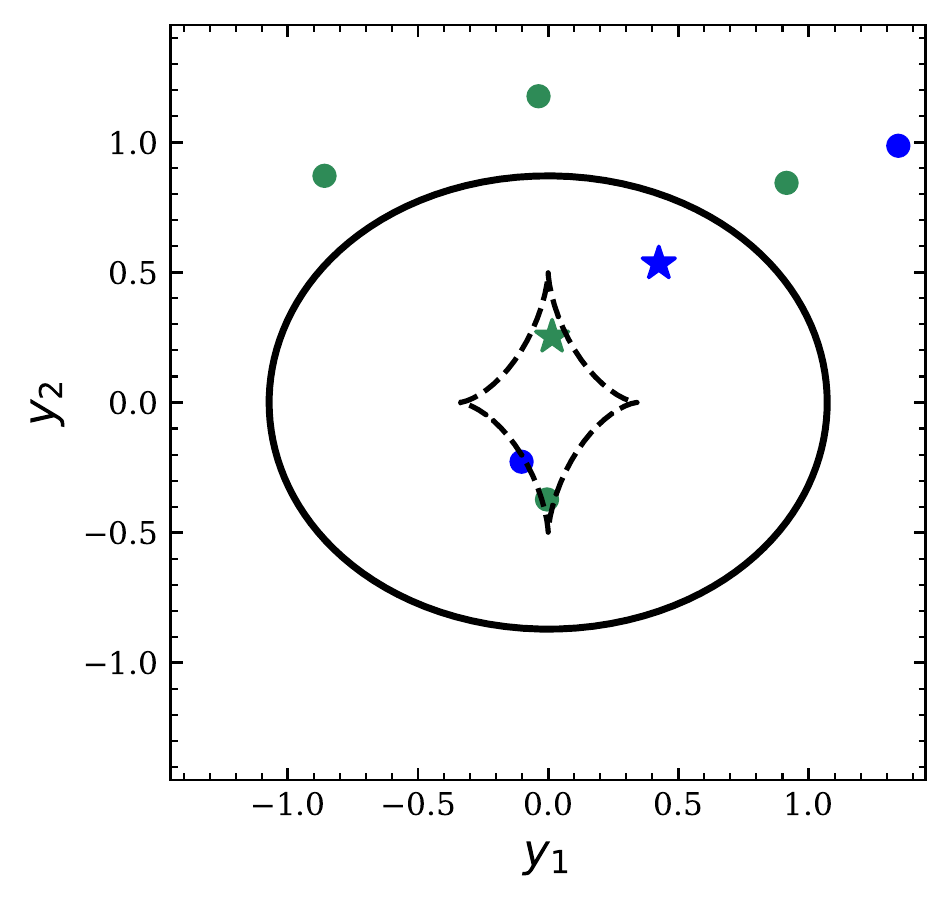}
	\caption{
An example illustrating the number of images produced based on the position of the source with respect to the caustics. When the source (green star) lies inside the inner caustic, four images (green circles) are produced. When the source (blue star) lies between the inner and outer caustics, two images (blue circles) are produced.}
		 \label{fig:example}
\end{figure}

There are two intrinsic properties of the SIE lens that determine its effective potential. These are the velocity dispersion $v$ (which can be thought of as a proxy for mass assuming the galaxy has virialized), and the axis ratio $q$. Lensing configurations with an SIE lens admit $2$ or $4$ images which depend on the position of the source with respect to the caustics (see Figure ~\ref{fig:example}). These are curves in the source plane where the magnification of the images formally diverge. If the source lies inside the inner caustic, four images (quad) will be generated. If the source lies between the inner and outer caustics, two images (double) will be produced.  \citep{kormann1994, koopmans2009}. 

\subsection{The time delay formula}

The Fermat potential $\phi$ is defined using the effective lens potential $\psi$ (cf. Eq.~\ref{eq:potential}) as \citep[e.g.,][]{schneider1992}:
\begin{equation}\label{eq:fermat_potential}
\phi(\vec{x}, \vec{y}) = |\vec{x} - \vec{y}|^2 - \psi(\vec{x})
\end{equation}
The time delay between two images A and B located at $\vec{x}_A,~\vec{x}_B$ is then acquired from the difference ($\Delta\phi$) in the Fermat potential evaluated at those positions:
\begin{equation}\label{eq:td}
c\Delta t = D_{\Delta t}\Delta\phi
\end{equation}
In the SIE model, the time-delay distance is given by \citep{kormann1994}:
\begin{equation}\label{eq:td_dist_sie}
D_{\Delta t} = (1 + z_l)X_0^2\frac{D_s}{D_l D_{ls}}
\end{equation}
where $X_0 = 4\pi (v^2/c^2) (D_{ls}/D_s)$ is a typical length scale associated with the lens \footnote{This length scale is the Einstein radius in the singular isothermal sphere (SIS) model.}. The SIE Fermat potential has a complicated form, and is given in Eq.~\ref{eq:potential_sie}.

\subsection{Leveraging lensing for early-warning}

We outline below two scenarios, involving lensed CBCs with EM counterparts, where some of these quantities can be acquired from direct measurements, while the others can be inferred assuming an SIE model for the lens. The first scenario requires at least two detectable images, while the second requires at least three.  We refer the reader to Appendix~\ref{app:td-from-locs} for details.

{\it Scenario 1:} The host galaxy of the CBC is identified from its EM counterpart \footnote{A concrete example of host-galaxy identification in real-time is provided by the EM-follow-up efforts of GW170817 \citep{GW170817-MMA}. The host galaxy was identified to be NGC4993 \citep{ngc4993}}. Additional image analysis reveals, at least, two images of the lensed host galaxy. Follow-up analyses further provide measurements of the image positions $\lbrace \vec{x}_i \rbrace$ of each of the images. Assuming SIE model for the lens, and assuming that the redshifts ($z_l$, $z_s$) are known, we can estimate $q, v, \vec{y}$ (Eqs.~\ref{eq:lens}, \ref{eq:potential}). Subsequently, we can predict the time-delays (cf. Eqs.~\ref{eq:td}, \ref{eq:td_dist_sie}, \ref{eq:potential_sie}) for the occurrence of the next CBC (and EM-counterpart) images. Note that such a scenario can happen either in a double configuration, or a quad configuration where only two of the four galaxy images can be confidently identified.

{\it Scenario 2:} Two images of a lensed CBC with EM counterparts have been observed, along with the positions of, at least, three images of the host galaxy. As with {\it Scenario 1}, two lensed images of the galaxy enable estimates of $q, v, \vec{y}$. Furthermore, in tandem with $q, v, \vec{y}$ and the measured positions of the lensed galaxy hosting the EM counterpart of the CBC, the time delay between the first two EM images of the CBC provide an estimate of the time-delay distance $D_{\Delta t}$ (cf. Eqs.~\ref{eq:td}, \ref{eq:td_dist_sie}, \ref{eq:potential_sie}). Using $D_{\Delta t}$, time-delays between other pairs of images can be straightforwardly estimated from the image positions and $q, v, \vec{y}$. This scenario has the advantage of not needing measurements of the source and lens redshifts, nor the assumption of a cosmology. Note that such a scenario can only happen in a quad configuration.

It would be germane to point out here that both scenarios considered assume that there is no offset between the positions of the CBC and its host galaxy. This may not be the case in general. Indeed, the position of GW170817 with respect to the center of its host galaxy, NGC4993, was found to be at an offset of $\sim 10''$ \citep{GW170817-MMA, Hjorth:2017yza}.  Nevertheless, within the SIE framework, this offset can be accounted for, provided it can be sufficently well measured. As with {\it Scenarios 1, 2}, the intrinsic lens parameters $q, v$ are acquired from the positions of the lensed images of the host galaxy. Next, from the coordinates of the observed EM images of the CBC and the inferred lens parameters, the position of the other EM images of the CBC ($\vec{x}_i$), and the unlensed position of the CBC ($\vec{y}$), can be calculated. The time delays between image pairs of the CBC are then evaluated using the available redshifts of the source and the lens. 

\begin{figure*}[htb]
	\centering 
	\includegraphics[width=0.9\linewidth]{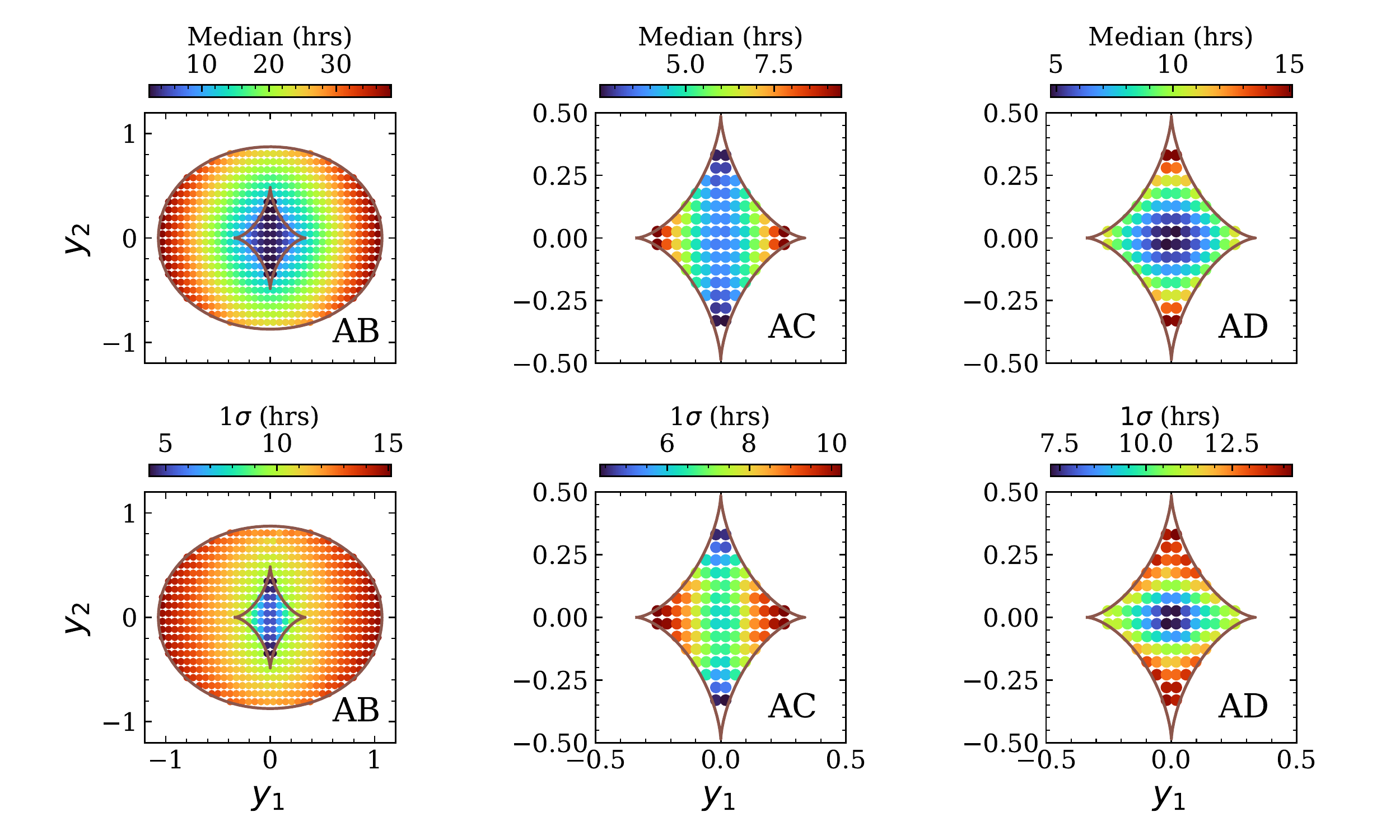}
	\caption{
Median (top) of the predicted time delays and the corresponding full-width of the $68\%$ confidence interval (bottom) shown for the double and quad configurations. The time delays of images (labelled as A, B, C, D, and ordered by arrival times) are computed with respect to the time of arrival of the first image (A). While both the median and confidence interval increase as the source moves away from the center, the uncertainties are mostly less than a day such that the relative error on the time delay spans $0.2 - 2$.}
		 \label{fig:error-plot} 
\end{figure*}

\subsection{Simulating time-delay distributions}

The early-warning times one might expect in an observing scenario are estimated by constructing realistic time-delay distributions of CBCs (BNS/NSBHs) that are both lensed and detectable. We account for the magnification of the GW images as well as the sensitivity of the GW detector network when deciding if an event is detectable. 

The lensing optical depth, evaluated from the parameter distributions of the lenses, determines the probability that GWs from a given CBC will encounter at least one lens in their journey towards the earth. This probability, in tandem with the intrinsic redshift distribution of the sources, determines the time delay distribution of the CBCs that will be lensed. However, not all lensed CBCs will be detectable. The detectability of the lensed CBC is determined by its extrinsic properties (distance, skylocation, inclination), its component masses ($m_1, m_2$), and the sensitivity of the GW detector network. Thus, the profile of the detectable time-delay distribution will also depend on these factors. We refer the reader to the Appendix~\ref{app:td-distribution} for details on the construction of this distribution for various observing scenarios. 
\vfill\null


%% file: results.tex
To study the effect of finite angular resolution of EM telescopes on the prediction of the time-delays, we first consider a typical lens system detectable (network S/N $\geq 8$) via GWs (from a BNS) in the O5 scenario. The values of the parameters of our lens model are chosen to be $q=0.54, v=111$ \rm{km s}$^{-1}$, $z_l=0.07, z_s=0.22$, corresponding to a lensed BNS detectable in O5. Keeping these lens model parameters fixed, we compute the time delay and image positions in the image plane by varying source positions on a grid spanning the source plane. The images are assumed to have an angular resolution of $0.05''$ at optical wavelengths, accessible to space-based telescopes, or ground based telescopes with adaptive optics technology.

\begin{figure}[htb]
	\centering 
	\includegraphics[width=1\linewidth]{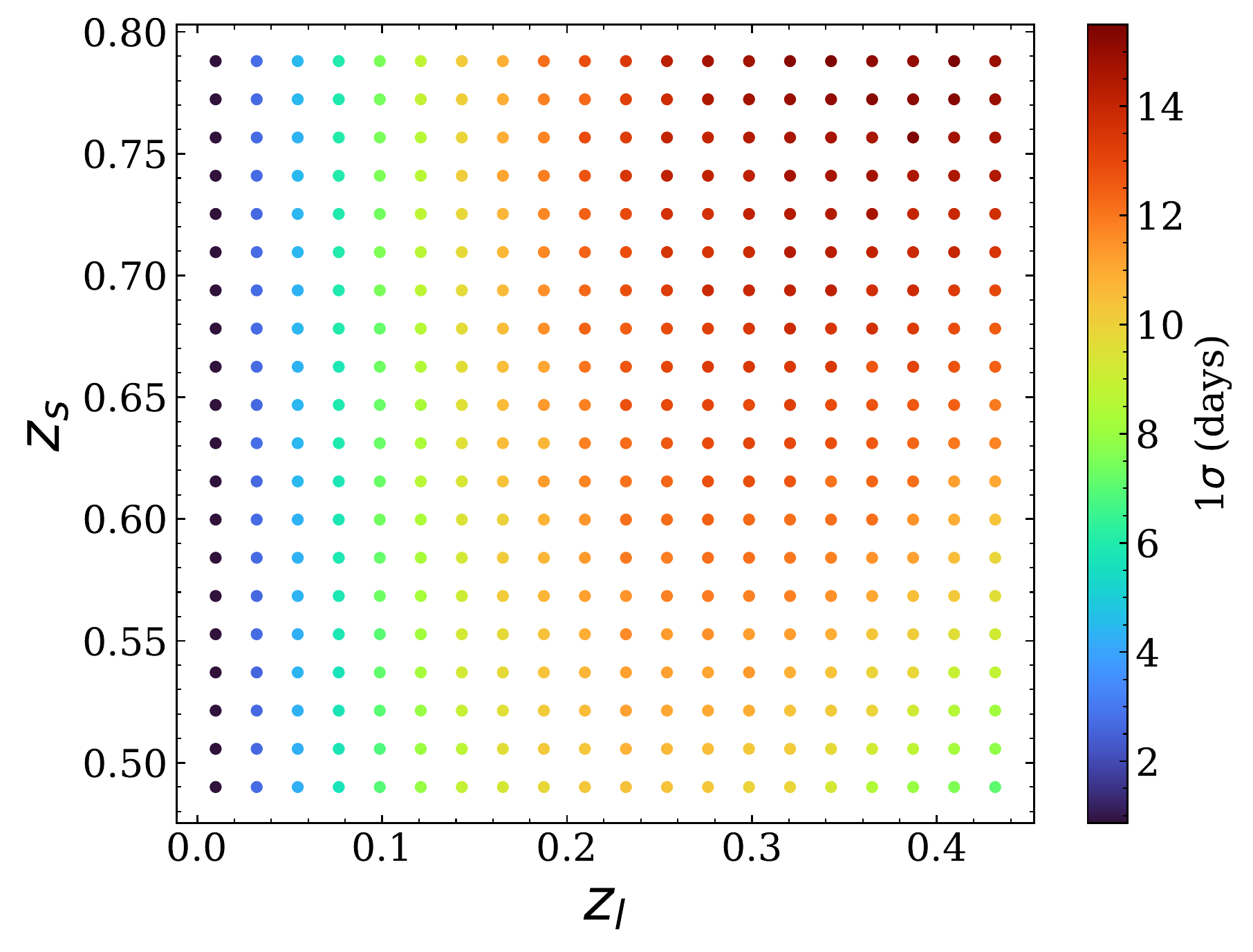}
	\caption{
The full-width of the $68\%$ confidence interval, pertaining to image-time delay predictions from host galaxy images. The time delays -- computed with respect to the time of arrival of the first detectable image -- and corresponding absolute width of the confidence interval, increase as both redshifts increase, although the relative error remains confined to $0.7 - 1.3$.}
		 \label{fig:error-plot-z}
\end{figure}

\begin{figure*}[htb]
	\centering 
	\includegraphics[width=0.45\linewidth, height = 0.34\linewidth]{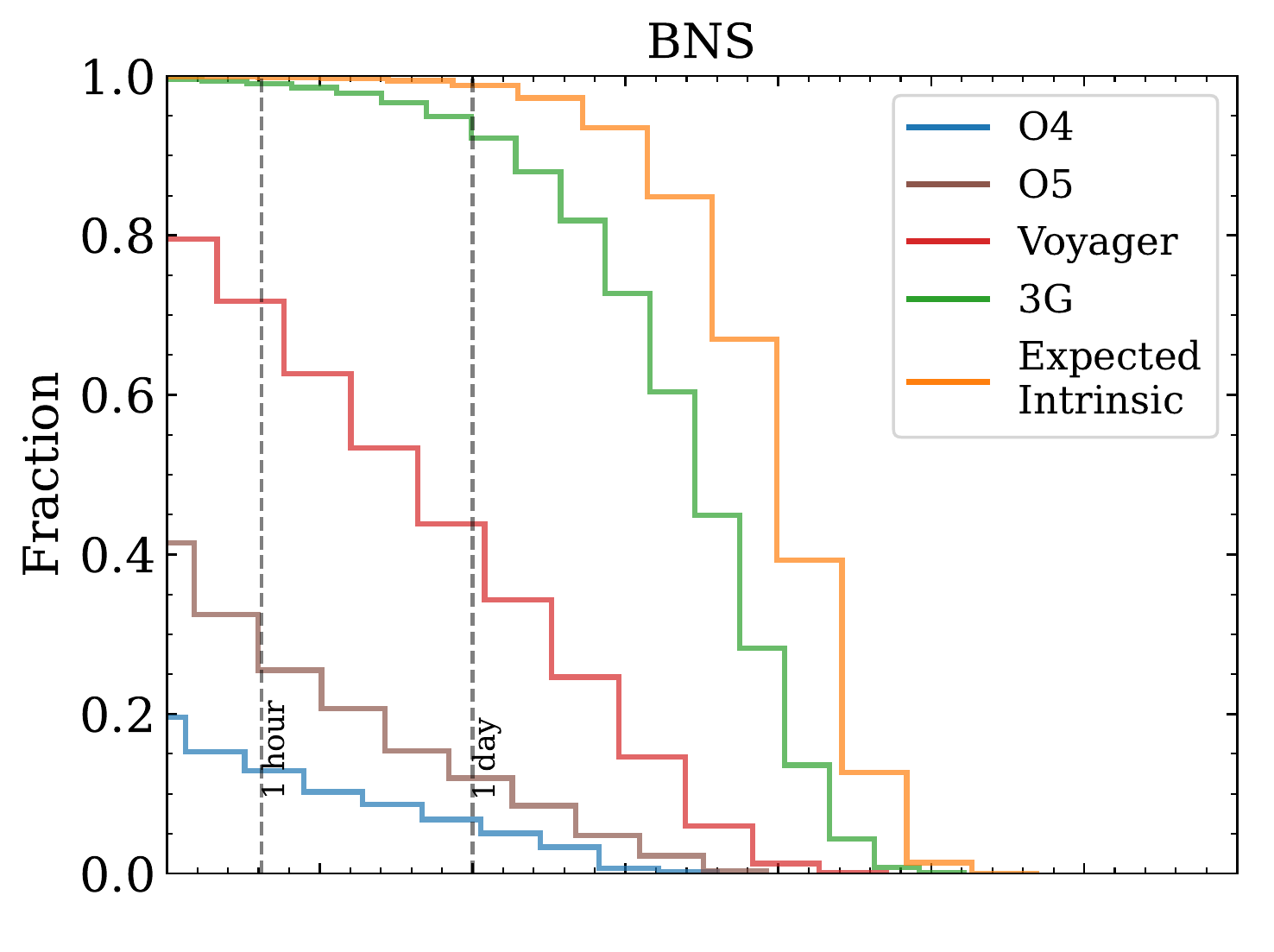}
    \includegraphics[width=0.45\linewidth, height = 0.34\linewidth]{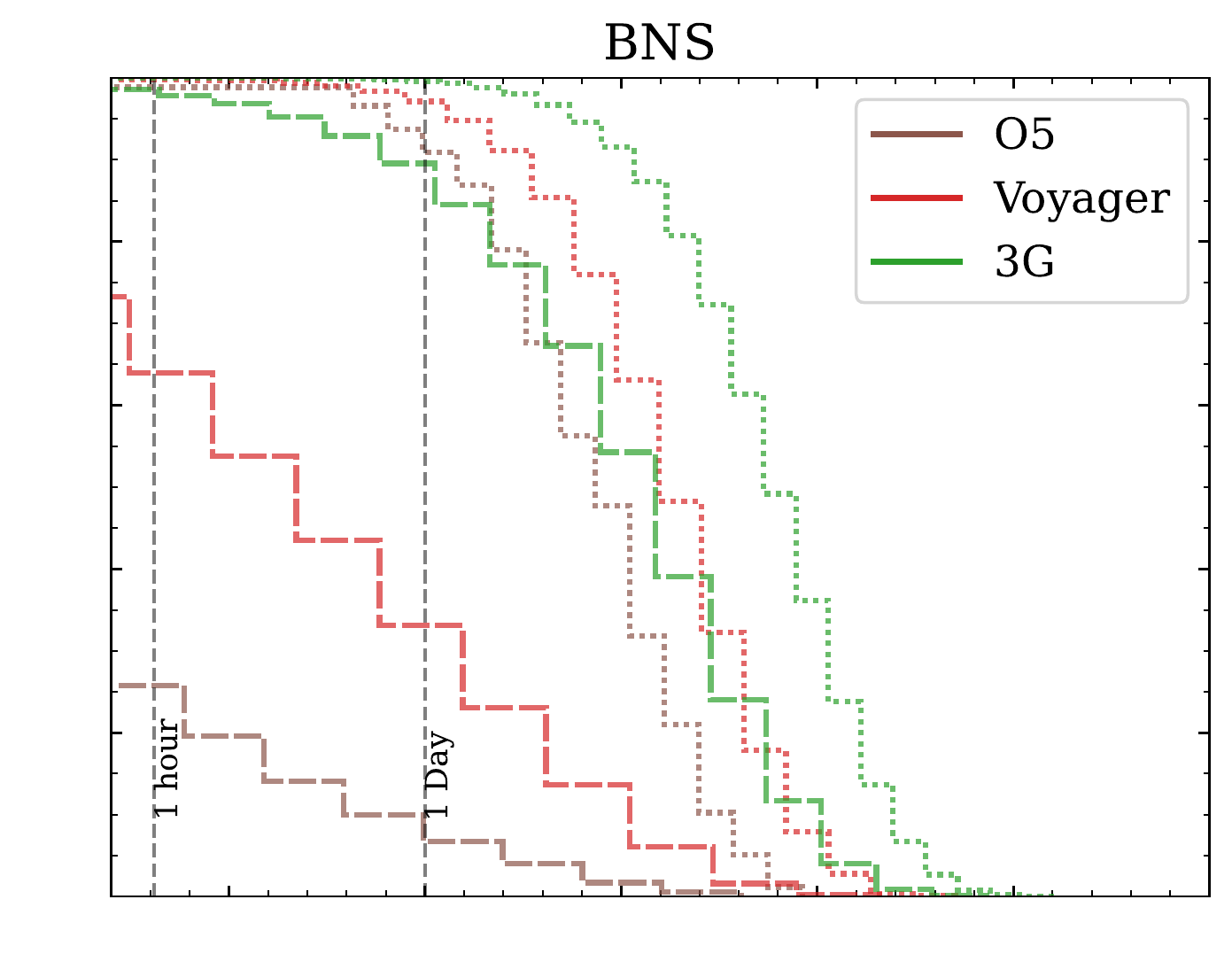}
   	\includegraphics[width=0.45\linewidth, height = 0.34\linewidth]{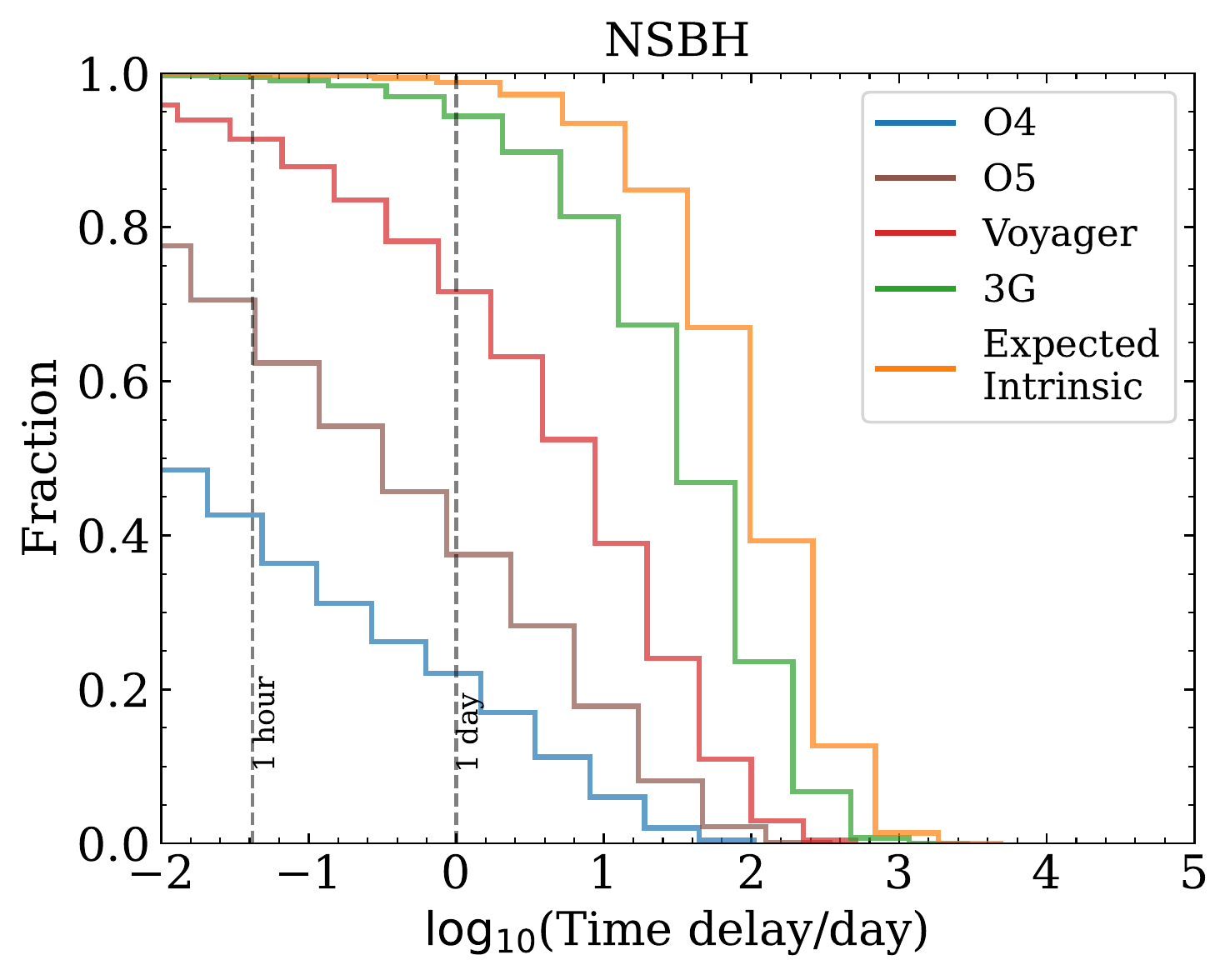}
    \includegraphics[width=0.45\linewidth, height = 0.34\linewidth]{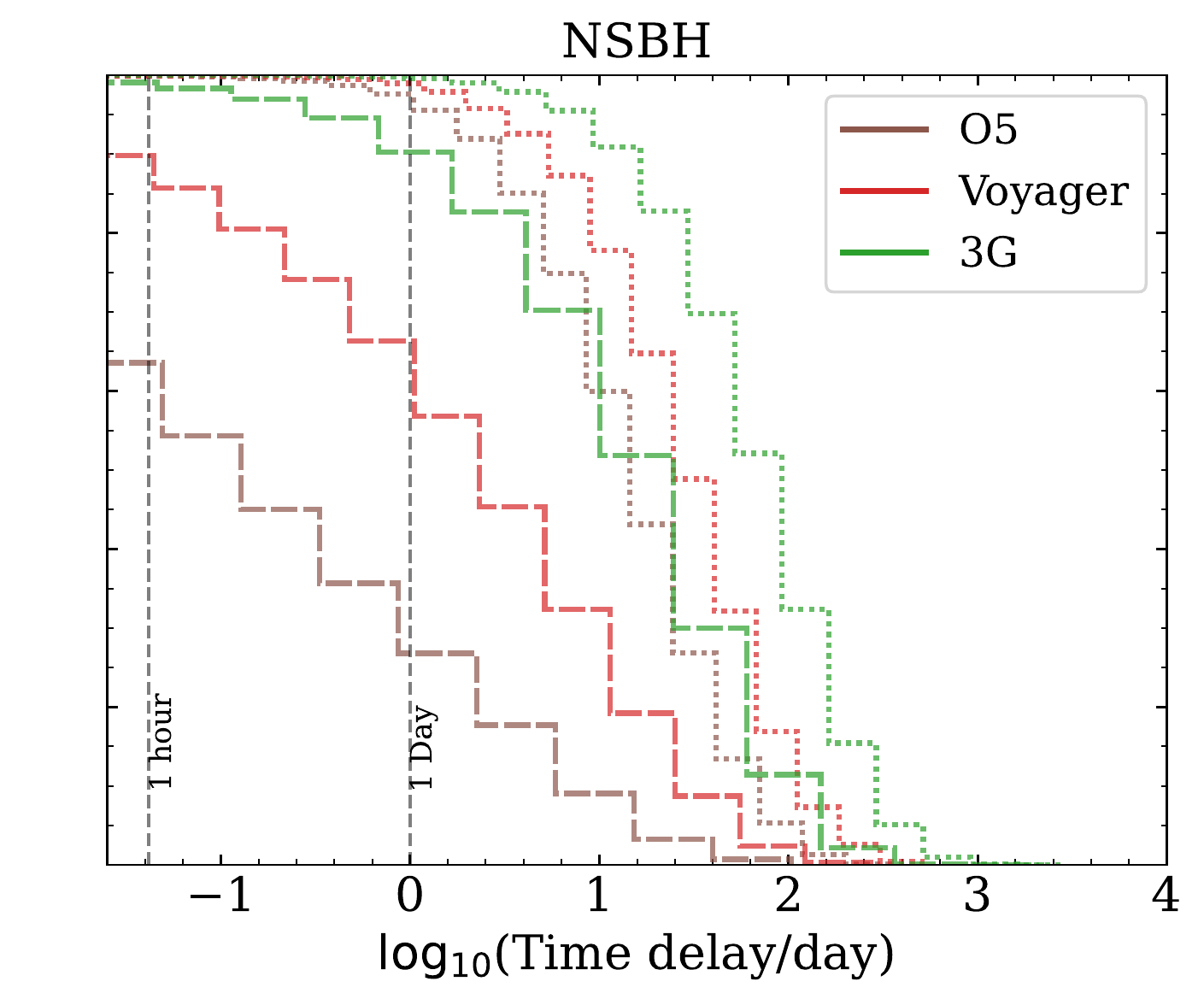}
	\caption{{\it Top left panel}: Time delay distributions (rendered as cumulative histograms) for detectable lensed BNS events in various observing scenarios. The histogram of the total number of lensed BNS events is also plotted.  As the sensitivity of the observing scenarios increase, the  distributions converge towards the histogram of the expected number of lensed events. Correspondingly, the fraction of large time delays also increases. {\it Top right panel}: The same distributions as on the top left but shown separately for doubles (dotted) and quads (dashed). The detectable quads have typically smaller time delays than the doubles. {\it Bottom left and right panels}: Same as the top left and right panels, except the sources are NSBHs. The detectable time delays are typically larger than the corresponding time delays for BNSs. This can be attributed to the larger horizon distance of GW detector networks for NSBHs compared to BNSs, and a resulting increased fraction of larger time-delays.}
		 \label{fig:BNS-td}
\end{figure*}

\begin{table*}
\centering
\begin{tabular}{|l|c|c|c|c|c|c|} \hline
Observing Run & Doubles $[\%]$ & Quads $[\%]$ & Quads (4 images) $[\%]$ & Quads (3 images) $[\%]$ & Quads (2 images) $[\%]$ \\\hline
O4 (BNS) & $3.30$ & $96.7$ & $0.36$ & $4.19$ & $92.2$ \\\hline
O4 (NSBH) & $10.8$ & $89.2$ & $ 1.65$ & $9.88$ & $77.7$ \\\hline
O5 (BNS) & $5.10$ & $94.9$ & $1.13$ & $6.65$ & $87.1$ \\\hline
O5 (NSBH) & $18.5$ & $81.5$ & $3.16$ & $14.4$ & $63.9$ \\\hline
Voyager (BNS) & $14.8$ & $85.2$ & $2.57$ & $14.7$ & $67.8$ \\\hline
Voyager (NSBH) & $37.0$ & $63.0$ & $6.23$ & $19.2$ & $37.7$ \\\hline
3G (BNS) & $60.8$ & $29.2$ & $7.42$ & $15.2$ & $16.6$ \\\hline
3G (NSBH) & $66.3$ & $33.7$ & $8.04$ & $13.5$ & $12.2$ \\\hline
\end{tabular}
\caption{Detectable lensed BNSs and NSBHs in the double and quad configurations, tabulated as percentages of the total detected number. The quads are further divided into categories based on the number of images within the quad that are detectable. The percentages suggest that if a lensed BNS or NSBH event is detected in O4 or O5, it will most likely be a quad detected as a pair consisting of the two loudest events in the quad. This is because, for the majority of the quads, the second and third images having higher magnifications than the first and the fourth images. As the sensitivity of the observing scenarios increase, the detected lensed events tend to have an increasing fraction of doubles. }
\label{tab:table}
\end{table*}


To estimate the errors on the time-delay predictions, we draw image coordinates $\tilde{x}_1, \tilde{x}_2$ \footnote{Angular image coordinate vector $\tilde{\vec{x}}$ acquired from observation, is a rescaled version of the position vector $\vec{x}$ that appears in the lens equation (Eq.~\ref{eq:lens}). See Appendix~\ref{app:td-from-locs} for details.} from a Gaussian centered on the true lensed-image position with a standard deviation of $0.05''$.  We similarly draw samples for source redshift $z_s$ assuming a relative error of $0.01$. 
We find that (see Figure~\ref{fig:error-plot}), for the system under consideration, the absolute errors are typically confined to within a day.  The relative errors for the majority of the positions span $0.2-2$. Time delays for doubles are typically larger than the time delays for quads, with some doubles having time delays up to $36$ hrs for the example we consider.

We also vary the source and lens redshifts, while keeping the source position and other parameters fixed, to study their effects on the precision of the predicted time delay (see Figure~\ref{fig:error-plot-z}). The values of the fixed parameters are $q=0.40, v=206$ \rm{km s}$^{-1}$. 

Higher source and lens redshifts result in larger time delays, and correspondingly larger error bars on their predictions assuming a fixed angular resolution. For sources out to redshifts of $\sim 1$ and lens redshifts of $\sim 0.5$, error-bars can exceed $2$ weeks for an angular resolution of $50$ milli-arcseconds. Nevertheless, the larger time delays of such lensing configurations could enable more extensive follow-up to improve on the error-estimates, either using higher-resolution telescopes and/or employing more sophisticated image analysis techniques. 

Using the prescription detailed in the Appendix \ref{app:td-distribution},  we determine the time delay distributions for lensed BNS and NSBH, for O4, O5, Voyager and the 3G scenario. We evaluate these distributions for detectable lensed event pairs only. A lens system is deemed detectable if at least two of its images have a network S/N $\geq 8$. The time-delays are calculated for all detectable images, setting the arrival time of the first detectable image in a given pair as the reference.  

The shape of these distributions is governed by two competing effects. Intrinsically, lensed BNSs and NSBHs tend to have larger $z_l, z_s$ and therefore larger time delays. On the other hand, the limited sensitivity of the detector network preferentially selects events with lower $z_l, z_s$.

The top left panel of Figure~\ref{fig:BNS-td} shows time-delay cumulative histograms for lensed BNSs detectable in various observing scenarios. In O4, $\sim 15\%~(5\%)$ of detectable lensed BNSs will have early-warning times of greater than $1$ \rm{hr}$~(1~\mathrm{day})$. This suggests that {\it if} a lensed BNS with a GW counterpart is detected in O4, then the subsequent image will likely occur well within an hour of the first image. This is consistent with corresponding estimates reported in \citep{smith2022} for BNSs. Thus, if EM-follow-up efforts identify the host galaxy images, then even an approximate conservative estimate of the upper limit on the time-delay would suffice. 
The percentages increase to $\sim 25\%~(10\%)$ in O5, $70\%~(45\%)$ in Voyager, and $\sim 100\%~(90\%)$ in 3G. Similarly, the bottom left panel shows time-delay cumulative histograms for lensed NSBHs detectable in various observing scenarios. In O4, $\sim 40\%~(20\%)$ of detectable lensed NSBHs will have early-warning times of greater than $1$ \rm{hr}$~(1~\mathrm{day})$. This increases to $\sim 60\%~(35\%)$ in O5, $90\%~(70\%)$ in Voyager, and $\sim 100\%~(95\%)$ in 3G.

We also evaluate the cumulative histograms of detectable quads and doubles separately, for both BNSs and NSBHs. These are depicted in the top right and bottom right panels of Figure~\ref{fig:BNS-td}. As expected, the time delays of detectable quads are significantly lower than the time delays of the detectable doubles. This is an observational bias arising from the second and third images, a spatially close pair with short time delays, having the highest magnifications among the four images of a quad \citep[see, e.g.][]{more2022}. 

Table~\ref{tab:table} shows the fraction of detected doubles, as well as quads detected as doubles, triples and quads, for both BNSs and NSBHs, across all observing runs considered in this work. We find that while doubles are intrinsically larger in number than quads, the detected doubles are fewer in number than the quads, for all observing scenarios except 3G. Nevertheless, since a large fraction of these quads are detected as doubles, {\it Scenario 2} will likely be rare, and may only come into effect in the Voyager and 3G scenarios.   

%% file: conclusion.tex

We describe below some of the caveats and challenges that we plan to address in follow-up work to make our method practicable. 

\paragraph{Detectability of the EM counterpart}

BNS and NSBH mergers are expected to produce EM counterparts spanning a wide range of frequencies. The counterparts include kilonovae (KNe), short gamma-ray bursts (sGRBs) and (potentially) precursors. Some of these are detectable to larger distances than others.  

KNe tend to become too faint to be observable beyond a few hundred Mpc. However, the lensed BNSs/NSBHs will be seen at much larger distances owing to the lensing magnification bias (see Figure~\ref{fig:redshift-distribution}). 
\cite{smith2022} suggest that lensed KNe (from BNSs) could be detected out to redshifts of $z \sim 2$ with the Vera Rubin Observatory \citep{LSST:2008ijt}. A tell-tale sign of the lensed nature of the event will be that the source-frame component masses inferred from the GW signal will appear to lie in the lower mass gap of the compact object mass spectrum.

Other counterparts such as sGRBs or precursors in the radio band, are typically detectable at cosmological distances. With magnification, these could be detectable to signficantly larger distances. 

As follow-up work, we propose to do a detailed study of the various EM-counterparts of lensed BNSs/NSBHs, and the distances out to which they can be observed.

\paragraph{Resolving galaxy images}

A crucial step in our method is the measurement of the angular separations between the multiply lensed host galaxy images which will likely involve a multi-step process.
Obviously, if the initial follow-up is being conducted with space-based telescopes such as the JWST \citep{JWST}, we will readily be able to extract the necessary lensing observables such as the image positions, relative magnifications, light profile of the host galaxy and the offset of the EM counterpart of the BNS/NSBH from its host galaxy, in the best case. 
In a more typical follow-up effort from the ground-based telescopes, the lens system will be imaged by metre-class telescopes, initially, for which the imaging quality tends be $\sim 1''$ or worse. Since the typical angular separations of lens systems tend to be about $1''$, we anticipate challenges in establishing the lensing nature in such cases. If lensing is confirmed, determining the lensed image observables accurately may still require higher-resolution imaging from either ground-based telescopes with adaptive optics, or space-based ones.

To reduce the latency of such a hierarchical process, it might be advantageous \footnote{The advantages of such an exercise are not limited to enhancing early-warning. See \citep{hannuksela2020} for additional benefits in the context of precision cosmology.} to catalog accurate predictions of time delays of strongly lensed images of galaxies produced in a homoegenous framework to minimise systematic biases arising from other modeling results in the literature. If lensed CBCs with EM counterparts are identified, with high probability, to belong to one of these lensed galaxies, predicting the occurrence of the next image could become trivial and very rapid.

As follow-up work, we plan to analyse existing galaxy-scale lens systems. Since the known lens systems are observed by telescopes of varying resolutions and imaging quality, we also plan to investigate their effects in estimating the corresponding error-bars on the predicted image time delays. This would help assess the amount of resolution required for our method to be realized in practice.

\paragraph{The lens model}

The expectation is that SIE adequately models galaxy lenses. However, more realistically, the density profiles of galaxies may depart from the isothermal assumption \citep{koopmans2009}, may possess a large external shear, or may have additional perturbers such as satellite galaxies. Such complexities may need to be accounted for in the model. We plan to inculcate some of these complexities and investigate their effects on the predicted time delays.


%% file: appendix.tex
\section{Estimation of the rate of detected and lensed BNS and NSBH mergers}\label{app:rate}

The rate estimation method follows Section 3.1 of \cite{LIGOScientific:2021izm}. We summarize the method here for the convenience of the reader: 

Computing the rate of lensed events requires assumptions on the model of the lens, the velocity dispersion function of the lens, as well as the redshift distribution of the sources (BNSs/NSBHs). The lens model is assumed to be SIS, and the dispersion function follows the Sloan Digital Sky Survey \citep{Choi:2006qg}. The evaluation of the rate of detectable lensed events additionally requires assumptions on the mass distribution of the CBC sources being lensed. This distribution is kept identical to what was used for the estimation of the time delay distribution (see Appendix~\ref{app:td-distribution}).

The governing equation for evaluating the rate of strongly lensed CBCs is given by:
\begin{equation}
R = \int \frac{dN}{dM_l}\frac{dD_c}{dz_l}\frac{R_s(z_s)}{1 + z_s}\frac{dV_c}{dz_s}\sigma(M_l, z_l, z_s, \rho, \rho_c)p(\rho\mid z_s)d\rho dz_s dz_l dM_h
\end{equation}
Here, $dN/dM_l$ is the mass spectrum of the lenses, $D_c, V_c$ are the comoving distance and volume, $\sigma$ is the lensing cross section, $\rho, \rho_c$ are the $S/N$ and threshold $S/N$ respectively, $p(\rho|z_s)$ is the $S/N$ distribution of the CBCs as a function of the source redshift, $R_s(z_s)$ is the redshift evolution of the merger rate, and $(1 + z_s)^{-1}$ accounts for cosmological time dilation.

We assume three models for $R_s(z_s)$, Madau-Dickinsion \citep{Madau:2014bja}, Oguri \citep{Oguri:2018muv}, and O3b \citep{LIGOScientific:2021psn}. The latter assumes the Madau-Dickinsion ansatz, but constrains the coefficients of the model with data from O1, O2 and O3. We evaluate the $90\%$ confidence interval from the distribution on $R$. 


\section{Image time delays from image locations: a derivation}\label{app:td-from-locs}

Let $\vec{X} = (X_1, X_2)$ be the physical position of the image in Cartesian coordinates in the lens plane, with respect to the optical axis. Similarly, let $\vec{Y} = (Y_1, Y_2)$ be the source location in the source plane. We define rescaled position vectors $\vec{x}, \vec{y}$ as:
\begin{equation}\label{eq:vec_rescale}
\vec{x} = \frac{\vec{X}}{X_0}, ~~~ \vec{y} = \frac{\vec{Y}}{Y_0}
\end{equation}
where $Y_0 = X_0 D_s/D_l$ (see below Eq.~\ref{eq:td_dist_sie}). The SIE lens equation in plane polar coordinates ($\vec{x} = (|\vec{x}|, \varphi)$) is given by \citep{kormann1994}:
\begin{equation}\label{eq:phi}
\left [y_1 + f_1(q, \varphi) \right]\sin\varphi - \left[ y_2 + f_2(q, \varphi) \right]\cos\varphi = 0
\end{equation}

where:

\begin{equation}\label{eq:funcs}
f_1(q, \varphi) \equiv \sqrt{\frac{q}{1-q^2}}\sinh^{-1}\left(\sqrt{\frac{1-q^2}{q^2}}\cos\varphi \right), ~~~ f_2(q, \varphi) \equiv \sqrt{\frac{q}{1-q^2}}\sin^{-1}\left(\sqrt{1-q^2}\sin\varphi \right)
\end{equation}
The solutions to this equation give the polar angle $\lbrace\varphi_i\rbrace$ of the images provided they satisfy the conditions \citep{haris2018}:
\begin{equation}\label{eq:conditions}
\varphi_i \in [0, 2\pi), ~~~ \left [y_1 + f_1(q, \varphi_i) \right]\cos\varphi_i - \left[ y_2 + f_2(q, \varphi_i) \right]\sin\varphi_i > 0
\end{equation} 
The cartesian coordinates of the image locations are then acquired by solving:
\begin{equation}\label{eq:im_coords}
x_{1,i} = y_1 + f_1(q, \varphi_i), ~~~ x_{2,i} = y_2 + f_2(q, \varphi_i)
\end{equation}

\subsection{Scenario 1}

For the problem at hand, the image locations are available from the angular coordinates of the host galaxy images, measured with respect to the optical axis. The task then is to acquire the source location and the lens parameters from measured coordinates. 

We therefore rewrite the above equation in terms of angular coordinates $\tilde{\vec{x}}_i = \vec{x}_i \tilde{X}_0$, where $\tilde{X}_0 \equiv X_0/D_l$:
\begin{equation}
\tilde{x}_{1,i} = y_1\tilde{X}_0 + \tilde{X}_0f_1(q, \varphi_i), ~~~ \tilde{x}_{2,i} = y_2\tilde{X}_0 + \tilde{X}_0 f_2(q, \varphi_i) 
\end{equation}
The angular image coordinates readily provide the polar coordinates: $\tan \varphi_i = x_{2,i}/x_{1,i} = \tilde{x}_{2,i}/\tilde{x}_{1,i}$. Inverting the tangent function gives more than one solution between $[0, 2\pi)$, with each solution lying in a different quadrant. $\varphi_i$ for each image is chosen such that it corresponds to the quadrant in which the image lies. The above pair of equations has four unknowns: $\vec{y}, q, \tilde{X}_0(q, v)$. Two image locations will provide two such pairs, resulting in a system of four equations with four unknowns which can be readily solved. 

Estimating the image time-delay from lens parameters, source and image locations, requires the SIE effective potential:
\begin{equation}\label{eq:potential_sie}
\psi_{\mathrm{SIE}}(\vec{x}) = \sqrt{\frac{q}{1-q^2}}|\vec{x}| \left[\sin\varphi\sin^{-1}\left(\sqrt{1-q^2}\sin\varphi \right) + \cos\varphi\sinh^{-1}\left(\sqrt{\frac{1-q^2}{q^2}}\cos\varphi\right) \right]
\end{equation}
From the effective potential, the Fermat potential at each of the image locations can be readily evaluated. The time delay between two images is then proportional to the difference in the corresponding Fermat potentials. The prefactor -- the time-delay distance $D_{\Delta t}$ -- requires estimates of the lens and source angular diameter distances ($D_l(z_l), D_s(z_s)$). These can be determined from measurements of the lens galaxy redshift ($z_l$) and the host galaxy redshift ($z_s$), assuming a cosmology.

\subsection{Scenario 2}

This scenario, which only works in a quad configuration where at least three images are detectable, acquires the time-delay distance $D_{\Delta t}$ from the time delay measurements of the first two CBC images, say $A, B$. Mathematically:
\begin{equation}
D_{\Delta t} = \frac{\Delta t_{AB}}{\Delta \phi (\vec{x}_A, \vec{x}_B, q, v, \vec{y})}
\end{equation}
where, as with {\it Scenario 1}, $\vec{x}_A, \vec{x}_B$ are acquired from observation of the galaxy images $A, B$, and $q, X_0(q, v), \vec{y}$ are acquired from any pair of the observed galaxy images.

The time-delay for another pair of images, say $C, D$, can then be straightforwardly computed from their measured locations as:
\begin{equation}
\Delta t_{CD} = D_{\Delta t}\Delta\phi(\vec{x}_C, \vec{x}_D, q, v, \vec{y})
\end{equation}
This scenario has the virtue of not needing estimates of $z_l, z_s$ and the cosmological parameters.

\section{Constructing time-delay distributions}\label{app:td-distribution}

We follow the prescription outlined in  \cite{haris2018} to construct the distribution of time-delays pertaining to detectable GW images of lensed BNSs and NSBHs, for second and third generation detector networks. 
\begin{figure*}[htb]
	\centering 
	\includegraphics[width=0.49\linewidth]{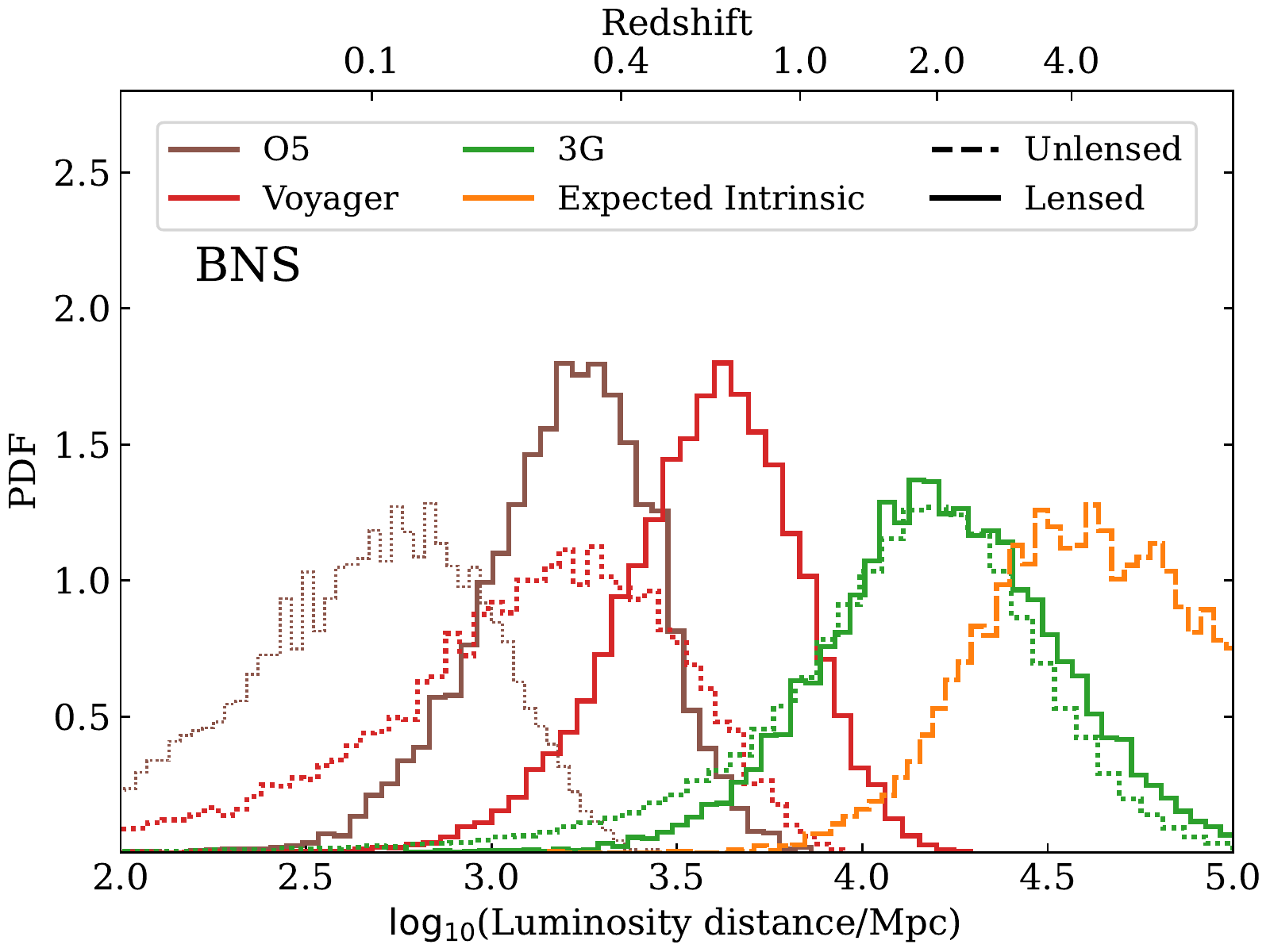}
    \includegraphics[width=0.49\linewidth]{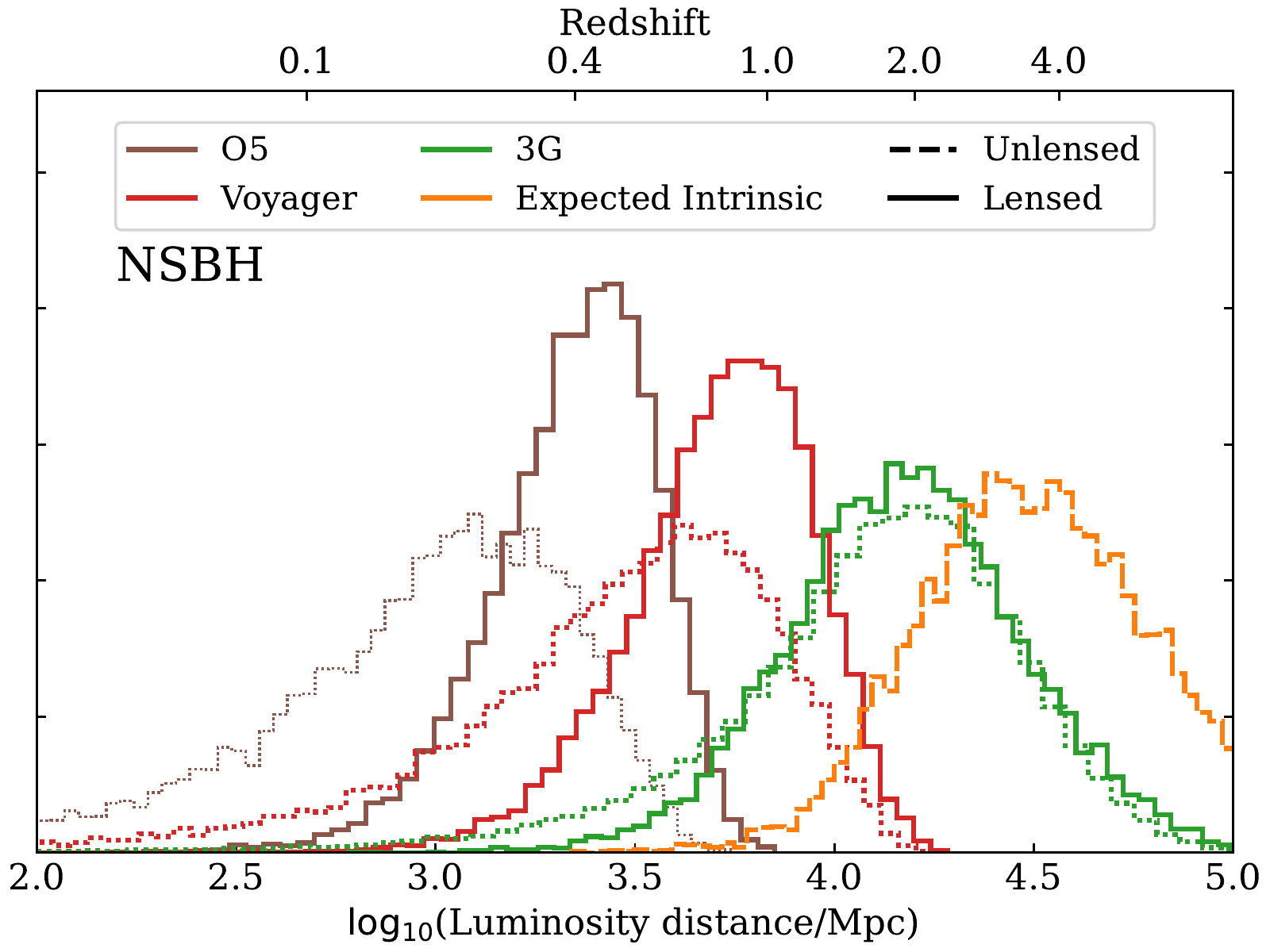}
    \caption{Simulated redshift distributions for lensed and unlensed BNSs ({\it left panel}) and NSBHs ({\it right panel}) detectable in various observing scenarios. We also include the expected intrinsic redshift distribution. }
    \label{fig:redshift-distribution}
\end{figure*}

\begin{enumerate}
\item {\it Lensed source redshift distribution:} We assume that the CBCs are distributed in redshift $z_s$ according to the Madau-Dickinson \citep{Madau:2014bja} distribution:
\begin{equation}
p(z) \propto (1 + z)^{\alpha}\left[1 + \left(\frac{1 +z}{1 + z_p} \right)^{\alpha + \beta}\right]^{-1}
\end{equation}
based on the star formation rate density history. We use the fiducial values of $\alpha = 2.7,~\beta = 2.9, z_p = 1.9$, which corresponds to a zero delay time between the formation of the stars in the binary and its merger.

Only a fraction of the CBCs will be lensed. This fraction depends on the so-called lensing optical depth $\tau(z_s)$, which is a measure of the probability that the GWs traveling from source to earth will encounter at least one lens. We estimate this probability assuming that the lenses have a constant number density and their velocity dispersion distribution follow early-type galaxies:
\begin{equation}
P(\mathcal{H}_L | z_s) = 1 - \exp(-\tau(z_s)), ~~~ \tau(z_s) = 4.17\times 10^{-6}\left(\frac{D^c(z_s)}{\mathrm{Gpc}}\right)^3
\end{equation}
We sample the Madau-Dickinson distribution  weighted by the lensing probability, $p(z_s, \mathcal{H}_L) = p(z_s)P(\mathcal{H}_L | z_s)$, to acquire the lensed source redshift samples (see Figure~\ref{fig:redshift-distribution}).

\item {\it Distribution of other extrinsic source parameters:} We assume that the sources are isotropically distributed
 in the sky, which amounts to drawing right ascension and declination samples uniformly in $\alpha \in [0, 2\pi), \cos\delta \in [0, 1]$. 
We also assume that the inclination -- the angle between the orbital angular momentum and the line of sight -- is isotropically distributed on the sphere: $\sin\iota \in [0, 1]$.

\item {\it Distribution of instrinsic source parameters:} The BNS component masses ($m_1, m_2$ measured in $M_{\odot}$) are drawn uniformly from $[1, 3]$, ensuring that $m_1 > m_2$. The NSBH primary mass is drawn uniformly from $m_1 \in [5, 95]$, and the secondary is drawn uniformly from $m_2 \in [1, 3]$. All binaries are assumed to be non-spinning. 

The conventional expectation is that NSBH systems with mass ratios greater than $\sim 6$ are not likely to produce a remnant mass post-merger. As a result, these are unlikely to produce EM counterparts. This is because the tidal forces on the NS outside the innermost stable circular orbit of the BH is not sufficient to disrupt it. Nevertheless, there are models that propose precursors in the radio band pre-merger for higher mass-ratio NSBH systems. We adopt an agnostic stance on the EM-Bright nature of NSBHs and do not place any restriction on their mass-ratio.

\item {\it Distribution of lens redshifts: } We define the ratio of comoving distances to the lens and the source as $x \equiv D^c(z_l)/D^c(z_s)$. We then draw this ratio from the polynomial:
\begin{equation}
p(x) = 30x^2(1-x)^2
\end{equation} 
The lens redshift samples are then acquired trivially from the ratio samples and the lensed source redshift samples drawn in step 1.

\item {\it Distribution of intrinsic lens parameters: } We borrow the ansatz used in \cite{collett2015} (which provides a fit using observed galaxy-galaxy lenses) to draw the velocity dispersion and the axis ratio samples. 

For the dispersion samples, the ansatz employs a generalized gamma distribution:
\begin{equation}
p(a) = a^{\alpha - 1}\exp(-a^{\beta})\frac{\beta}{\Gamma(\alpha/\beta)}
\end{equation}
with $\alpha = 2.32,~\beta = 2.67$. The dispersion sample is then acquired from a sample $a$ as $v = a\sigma_{\star}$, with $\sigma_{\star} = 161 \mathrm{km}/\mathrm{s}$.

Given $a$, a sample $b$ is drawn from a Rayleigh distribution repeatedly until one satisfying $b < 0.8$ is found:
\begin{equation}
p(b) = \frac{b}{s^2}\exp\left(-\frac{b^2}{2s^2}\right), ~~~ b \in [0, \infty), ~~~ s = 0.38 + 0.09177a
\end{equation}
The axis ratio sample is then calculated as $q = 1 - b$.

\item {\it Distribution of sources in the source plane: } The sources are distributed uniformly in the source plane. Only those that lie within the outer caustic (which ensures the production of multiple images) are kept. This is equivalent to drawing $y_1, y_2$ from $p(y) \propto y$, with:
\begin{equation}
y_1 \in \left(0, \sqrt{\frac{q}{1 - q^2}}\cosh^{-1}\left [\frac{1}{q} \right]\right)
\end{equation}

\[
    y_2 \in 
\begin{cases}
    \sqrt{\frac{q}{1 - q^2}}\cos^{-1}[q],& \text{if } q >  q_0 \\
     \sqrt{\frac{1}{q}} - \sqrt{\frac{q}{1 - q^2}}\cos^{-1}[q],              & \text{if } q < q_0
\end{cases}
\]
where $q_0 = 0.3942$ is the solution to $2q_0\cos^{-1}q_0 - \sqrt{1 - q_0} = 0$.

For $q < q_0$, the inner caustic intersects the outer caustic, and the vertical cusps of the former lie outside the latter. In such cases, it is possible to have multiple images when the source lies outside the outer caustic but inside the inner caustic. To account for such cases, we always ensure that the plane being sampled encompasses both caustics, and solve the lens equation repeatedly for each draw of $y_1, y_2$, and keep only those samples that produce multiple images.

\item {\it Detectable images and their time delays: } The detectability of an image depends on its (de)magnified amplitude and the sensitivity of the detector network. Steps 1 through 6 enable us to solve the lens equation to generate samples of the image coordinates $\vec{x}_i$ in the lens plane, following Eqs.~\ref{eq:vec_rescale}, \ref{eq:phi}, \ref{eq:funcs}, \ref{eq:conditions}, \ref{eq:im_coords}. We evaluate the magnification $\mu$ of each of the images using:
\begin{equation}
\frac{1}{\mu} = 1 - \sqrt{\frac{q}{x_1^2 + q^2 x_2^2}}
\end{equation}
The GW amplitude of each image is given by $h_i = \sqrt{\mu_i}h$, where $h$ is the unlensed amplitude after accounting for the detector's response to each polarization via the antenna pattern functions. The time delay of the images also incurs an additional phase factor, but this has no contribution \footnote{This in general is not true for systems where higher harmonics have a non-trivial contribution to the total GW amplitude, and the image is of Type-II. We do not account for this effect in this work, and only use the dominant GW mode.} to the optimal signal-to-noise ratio (SNR) used to assess the detectability of the image \footnote{The values of the antenna pattern functions of the detectors could be non-trivially different for different images, especially for large time delays. The time-delays have therefore been accounted for there.}. 

A lensed configuration is said to be detectable if at least two images have an optimal network S/N $\geq 8$, where the S/N of the ith image is evaluated as:
\begin{equation}
\left(\mathrm{S/N}\right)_i = 4\mathrm{Re}\int_0^{\infty}\frac{|h_i(f)|^2}{S_n(f)}df
\end{equation}
and $S_n(f)$ is the noise power spectral density (PSD) of the detector. The network S/N is the optimal S/N of each detector in the network added in quadrature. The time delays are evaluated for all detectable images (using the arrival time of the first detectable image as reference), from Eqs.~\ref{eq:fermat_potential}, \ref{eq:td}, \ref{eq:td_dist_sie}, \ref{eq:potential_sie}.

\item {\it Observing Scenarios:} We define the detector networks for the different scenarios considered in this work. See \citep{KAGRA:2013rdx, Hall2019, CE, ET} for additional details on the observing scenarios, and \citep{psds, pycbc} for details on the assumed noise PSDs. 
\begin{itemize}
\item {\it O4:} 2 Advanced LIGO detectors, 1 Virgo detector, 1 KAGRA detector.
\item {\it O5:} 3 LIGO detectors, including LIGO India, at A+ sensitivity, 1 Virgo and 1 Kagra detector, at target sensitivity.
\item {\it Voyager:} 3 LIGO detectors at Voyager sensitivity, 1 Virgo and 1 Kagra detector at target sensitivity.
\item {\it 3G:} 2 Cosmic Explorers located at Hanford and Livingston, and 1 Einstein Telescope at Virgo's current location.

\end{itemize}
\end{enumerate}